\documentclass[]{spiejour}  
\usepackage[]{graphicx}
\usepackage{amsmath}
\usepackage{amssymb}
\usepackage[]{graphicx}
\usepackage{epsfig}

\def\##1{{\bf #1}}
\def\~#1{{\underline {\mathcal#1}}}
\def\+#1{{{\mathcal #1}}}
\def\=#1{\underline{\underline #1}}

\def\.{\mbox{ \tiny{$^\bullet$} }}

\title{Theoretical aspects of quantum state transfer, correlation measurement and electron-nuclei coupled dynamics in quantum dots} 

\author{Toshihide Takagahara and \"Ozg\"ur \c{C}akir}

\affiliation{Department of Electronics and Information Science,
Kyoto Institute of Technology,
Matsugasaki, Kyoto 606-8585, JAPAN\\
\linkable{takaghra@kit.ac.jp} \\
and \\
CREST, Japan Science and Technology Agency, 4-1-8 Honcho, Kawaguchi, 
Saitama 332-0012, JAPAN}

\begin{document} 
  \maketitle 

\begin{abstract}
Photons and electrons are the key quantum media for the quantum information processing
based on solid state devices.  The essential ingredients to accomplish the quantum
repeater were investigated and their underlying physics were revealed.  The relevant elementary processes of the quantum state transfer between a single photon and a single electron were analyzed,
to clarify the conditions to be satisfied to achieve the high fidelity of the quantum state transfer. An optical method based on the Faraday rotation was proposed to carry out the Bell measurement of two electrons which is a key operation in the entanglement swapping for the quantum repeater and its feasibility was confirmed.
Also investigated was the quantum dynamics in the electron-nuclei
coupled spin system in quantum dots and a couple of new phenomena were predicted related
to the correlations induced by the hyperfine interaction, namely, bunching and revival in the electron spin
measurements. These findings will pave the way to accomplish the efficient and robust quantum repeater and nuclear spin quantum memory. 
\end{abstract}


\section{INTRODUCTION}
\label{sect:intro}  
 Coherent manipulation of quantum states is a critical step toward many
novel  technological applications ranging from manipulation of qubits in
quantum logic gates\cite{ben,bar,bio,tro,che} to controlling the reaction pathways of
molecules.  In the field of the quantum state control by optical means, both
Rabi oscillation and quantum interference play the central roles. The
exciton Rabi splitting was observed in the luminescence spectrum of a
single InGaAs quantum dot and the exciton Rabi oscillation was also
observed in the spectroscopy of a single GaAs or InGaAs quantum dot
\cite{sti,kam,hto,zre,tak,wan,wan1}. The two-qubit CROT (controlled rotation) gate operation was
demonstrated using two orthogonally polarized exciton states and a
biexciton state in a GaAs quantum dot\cite{xia}.  Unfortunately, however,
the decoherence/dephasing times of excitons and biexcitons in these quantum
dots are limited by the radiative lifetimes($\sim$ 1 ns) even at low
temperatures. Thus a qubit with a longer decoherence time is desirable for
the application to the quantum information processing.
Electron spins in semiconductor quantum dots (QDs) are considered as one of the
most promising candidates of the building blocks for quantum information
processing\cite{los,ima}  due to their robustness
against decoherence effects\cite{gol,sem}.
In double QD systems, initialization and coherent manipulation of the electron spin have been realized, with 
coherence times extending to $1~\mu$s\cite{pet,kop}.

Recently, a quantum media converter
from a photon qubit to an electron spin qubit was proposed for quantum
repeaters\cite{yab,kos1}. Quantum information can take several different
forms and it is preferable to be able to convert among different forms. 
One form is the photon polarization and another is the electron spin
polarization. Photons are the most convenient medium for sharing quantum
information between distant locations.  However, it is necessary to realize
a quantum repeater in order to send the information securely over a very
long distance overcoming the photon loss. A quantum repeater requires two essential ingredients, namely, the
quantum state transfer between a photon and an electron spin and the correlation (Bell) measurement between two 
electrons created by the quantum state transfer from two different photons. Additionally it is desirable to 
have a long-lived quantum memory based on the nuclear spins. For the quantum state transfer,
 a strained InGaAs/InP quantum dot was proposed
as a preferable device based on the $g$-factor engineering\cite{kos2}.  In the actual
operation, the photoexcited holes are to be quickly swept out of the quantum
dot to project the photon polarization onto the electron spin polarization,
preserving the quantum coherence. We have analyzed the performance of this
operation and clarified the conditions to be satisfied to achieve a high value of the purity of the transferred quantum state or the fidelity of the quantum state transfer.

 Another fundamental element for realizing the quantum repeater is the Bell (quantum correlation)  measurement for the entanglement swapping. In our case this measurement is carried out for two electrons which are created by the quantum state transfer from two separate photons. Here we propose an optical method to do this Bell measurement based on the Faraday (in the transmission geometry) or Kerr (in the reflection geometry) rotation and estimate the feasibility.

 The electron spin decoherence time reported so far for low temperatures is about a few microsecond and
is not sufficiently long for the secure
quantum information processing. Thus the nuclear spin quantum memory will be eventually required and the
robust quantum state transfer should be realized between the electron spin and the nuclear spins in order 
to store and retrieve the quantum information. For that purpose, the fundamental features of dynamics 
in the electron-nuclei coupled system should be investigated. Here we reveal a sequence of back-actions
between the electron spin and the nuclear spins through the quantum state measurements and predict a couple of new phenomena. These findings will
open the way to realize the quantum state purification of nuclear spins, elongation of the electron spin decoherence time
and the nuclear spin quantum memory.

\section{QUANTUM STATE TRANSFER}
In the quantum state transfer between the photon qubit and the electron qubit, the one-to-one
correspondence should be established between the photon polarization and the electron spin 
polarization. In other words, the one-to-one correspondence between the Poincare sphere for a photon and the Bloch sphere for an electron spin should be realized as perfectly as possible. Since the photon energy is independent of the polarization direction, the electron energy should also be independent of the spin direction. This can be realized by the $g$-factor engineering\cite{kos2}.  The light-hole exciton is preferable than the heavy-hole 
exciton because of the characteristic optical selection rule and the possible Zeeman splitting 
in the in-plane magnetic field. In order to have the light-hole states as the ground hole states,
the strained quantum well (QW) structure is necessary and can be realized in InGaAs/InP QW structures. 
In these strained QW structures the light-hole states are written as\cite{vri}
\begin{eqnarray}
&&|\frac{3}{2}\frac{1}{2}\rangle=\sqrt{\frac{2}{3}}|10\rangle |\alpha \rangle +\sqrt{\frac{1}{3}}|11\rangle |\beta\rangle \;,\\
&&|\frac{3}{2}-\frac{1}{2}\rangle=\sqrt{\frac{1}{3}}|1-1\rangle |\alpha\rangle +\sqrt{\frac{2}{3}}|10\rangle |\beta\rangle\;,
\end{eqnarray}
where the usual angular momentum representation $|j, m\rangle$ is employed and $|10\rangle |\alpha\rangle$ indicates a $p_z$-like orbital with the up-spin, for example. Then the
relevant Hamiltonian for the light-hole states under an in-plane magnetic field along the $x$ axis are 
represented by
\begin{equation}
-g_{\ell h} \mu_B B S_x= - \frac{2}{3} g_{\ell h} \mu_B B \left( \begin{array}{cc}
0 & 1 \\
1 & 0 \end{array} \right) \;,
\end{equation}
where $g_{\ell h}$ is the g-factor of the light-hole and the eigenstates are given by
\begin{equation}
|\ell h_1\rangle=\frac{1}{\sqrt{2}}\left(|\frac{3}{2}\frac{1}{2}\rangle+|\frac{3}{2}-\frac{1}{2}\rangle \right)\;, 
\quad |\ell h_2\rangle=\frac{1}{\sqrt{2}}\left(|\frac{3}{2}\frac{1}{2}\rangle - |\frac{3}{2}-\frac{1}{2}\rangle
\right)\;.
\end{equation}
The selection rules of the optical transition can be derived by taking into account the conservation
of the orbital angular momentum and the decomposition of the linear polarization into the
circular polarization:
\begin{eqnarray}
&& \hat{x} \cos \omega t= \frac{1}{2}(\hat{x}+i\hat{y}) \cos \omega t + 
\frac{1}{2}(\hat{x}-i\hat{y}) \cos \omega t \;,\\
&& \hat{y} \cos \omega t= \frac{1}{2i}(\hat{x}+i\hat{y}) \cos \omega t - 
\frac{1}{2i}(\hat{x}-i\hat{y}) \cos \omega t \;,
\end{eqnarray}
where $\hat{x}\;(\hat{y})$ denotes the unit vector in the $x\;(y)$-direction.
Assuming $|\ell h_1\rangle$ to be the ground hole state (namely, $g_{\ell h} > 0$), we have the optical
selection rules:
\begin{equation}
|\ell h \rangle_1 \longrightarrow \frac{1}{\sqrt{2}} (|c \uparrow \rangle - | c \downarrow \rangle )=-\; 
|e_1 \rangle 
\end{equation}
for the $x-$polarized light and
\begin{equation}
|\ell h \rangle_1 \longrightarrow \frac{-i}{\sqrt{2}} (|c \uparrow \rangle + | c \downarrow \rangle )=-i\; 
|e_2\rangle
\end{equation}
for the $y-$polarized light, respectively, where $|c \uparrow(\downarrow) \rangle$ represents the conduction band electron state with the up (down) spin in the $z$ direction and $|e_1\rangle(|e_2\rangle)$ is the electron eigenstate with the spin aligned
 in the $-x\;(x)$ direction. The schematic energy levels are plotted in Fig. 1.

\begin{figure}[!t]
\begin{center}
\includegraphics[width=8cm]{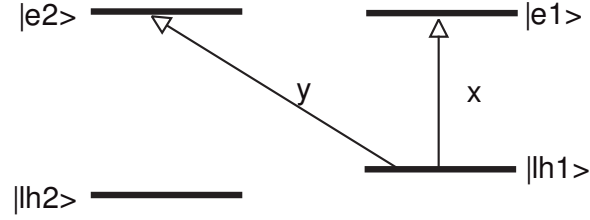}
\end{center}
\caption{{\fontsize{9pt}{9pt}\selectfont Schematic energy levels in a strained quantum well under an in-plane ($|| x$) magnetic field.}}
\end{figure}

 According to these selection rules, we can excite a linear combination of exciton states which have
a common hole state. When a linearly polarized light comes in with polarizaion given by
\begin{equation}
\cos \theta \; \hat{x} + \sin \theta \; \hat{y} \;, \label{eqq5}
\end{equation}
the excited state can be written as
\begin{equation}
 -\cos \theta |e_1\rangle |\ell h_1\rangle - i \sin \theta |e_2\rangle |\ell h_1\rangle 
=- \cos \theta |e_1, h_1\rangle -i \sin \theta |e_2, h_1\rangle \;, \label{eqq1}
\end{equation}
where the direct product state $|e_1\rangle |\ell h_1\rangle$ is simply denoted by $|e_1, h_1\rangle$ 
for example and
 the representation in the form of the electron-hole pair is used instead of the exciton representation. It is 
beyond the scope of this article to discuss fully the exciton effect in the quantum state transfer.
Now we study influences of relaxation processes of the electron and the hole on the quantum state 
transfer. In the case of a gate-controlled quantum dot, the electron is three-dimensionally confined,
whereas the hole is not confined and is absorbed into the negatively biased gate, leaving behind only the 
photoexcited electron.  Thus the important issue is the quantum nature of the electron state after 
extraction of the hole: How much is the electron spin coherence retained after the process ? 
To answer this question, we will analyze the time evolution of the whole system based on the density
matrix formalism. Just after the photoexcitation given by Eq. (\ref{eqq1}), the density matrix of the electron-hole system is supposed to be given by
\begin{eqnarray}
&&\rho(t=0)=\cos^2 \theta |e_1,h_1\rangle\langle e_1,h_1| + \sin^2 \theta |e_2,h_1\rangle\langle e_2,h_1|
\nonumber \\
&& \hspace{1.3cm} +i \sin \theta \cos \theta |e_2,h_1\rangle\langle e_1,h_1|-i\sin \theta \cos \theta |e_1,h_1\rangle\langle e_2,h_1|\;.
\label{eqq4}
\end{eqnarray}
Then the hole is extracted into the negatively biased gate electrode. In order to simulate this process, we introduce
a model consisting of three hole states as depicted in Fig. 2; one of them is the hole state 
$|h_1 \rangle$ created in a quantum dot by 
the photoexcitation, the second one is an intermediate hole state $|h_2 \rangle$ representing a 
delocalized state around the gate electrode and the third one is the swept-out state in the gate. The most
important mechanism degrading the electron spin coherence is the electron-hole exchange interaction which 
induces the spin state mixing.  During the hole extraction to the negatively biased electrode the hole spin 
relaxation occurs 
and the electron spin states are mixed up, leading to the degradation of quantum state transfer.
\begin{figure}[!t]
\begin{center}
\includegraphics[width=8cm]{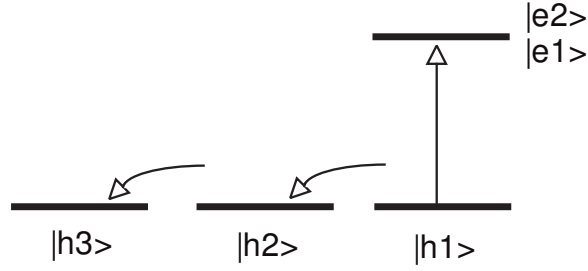}
\end{center}
\caption{{\fontsize{9pt}{9pt}\selectfont Schematic relaxation paths in the quantum state transfer from a photon to an electron spin. Curved
arrows indicate the transfer (tunneling) process of the photoexcited hole. $|h3\rangle$ represents symbolically the final destination of the extracted hole.}}
\end{figure}

\subsection{Dynamics of quantum state transfer}

 To set up the equations of motion for the density matrix, we take into account six basis states composed of 
direct products of two electron states $|e_1\rangle$ and $|e_2\rangle$ and three hole states $|h_1\rangle \;, 
|h_2\rangle$ and $|h_3\rangle$:
\begin{equation}
|e_1, h_1\rangle, |e_2, h_1\rangle, |e_1, h_2\rangle, |e_2, h_2\rangle, |e_1, h_3\rangle, |e_2, h_3\rangle\;.
\end{equation}
In this representation, the equation of motion for the density matrix is given by
\begin{eqnarray}
&&\dot{\rho} = -\frac{i}{\hbar}[H, \rho] + \Gamma \rho \;, \label{eqq2} \\
&&H= H_0 + H_{{\rm exch.}} \;,\\
&&H_0= E_e (|e_1 \rangle \langle e_1| + |e_2 \rangle \langle e_2|) - E_{h_1} |h_1 \rangle
\langle h_1| - E_{h_2} |h_2 \rangle\langle h_2| - E_{h_3} |h_3 \rangle\langle h_3| \;, \\
&& \Gamma \rho = t_{12} |h_2\rangle \langle h_1|\rho|h_1 \rangle \langle h_2| +t_{23}
|h_3\rangle \langle h_2|\rho|h_2 \rangle \langle h_3|  
 -t_{12} |h_1\rangle \langle h_1|\rho|h_1 \rangle \langle h_1|  \nonumber \\
&&-t_{23}|h_2\rangle \langle h_2|\rho|h_2 \rangle \langle h_2|  - \sum_{i\neq j} \gamma^h_{ij}\{|h_i \rangle\langle h_i|\rho|h_j\rangle\langle h_j|+
|h_j \rangle\langle h_j|\rho|h_i\rangle\langle h_i| \} \nonumber \\
&& +\gamma^e_1 |e_2\rangle \langle e_1|\rho|e_1 \rangle \langle e_2| -\gamma^e_1 |e_1\rangle
\langle e_1|\rho|e_1 \rangle \langle e_1| \nonumber \\
&&- \gamma^e_{12}\{ |e_1\rangle \langle
e_1|\rho|e_2 \rangle \langle e_2|+ |e_2\rangle \langle e_2|\rho|e_1 \rangle \langle e_1|\} \;,
\end{eqnarray}
where $t_{ij}$ is the transfer (tunneling) rate from the hole state $|h_i\rangle$ to $|h_j\rangle$, 
$\gamma^e_1$  the
electron spin relaxation rate from the electron state $|e_1\rangle$ to $|e_2\rangle$, $\gamma^e_{12}$
the electron spin decoherence rate and $\gamma^h_{ij}$ is the dephasing rate of the hole state coherence between $|h_i\rangle$ and $|h_j\rangle$.  
Concerning the electron-hole exchange interaction we employ the following Hamiltonian:
\begin{equation}
H_{{\rm exch.}} = W \cdot \left( \begin{array}{ccccccc}  & |e_1 h_1 \rangle & |e_2 h_1
\rangle & |e_1 h_2 \rangle & |e_2 h_2 \rangle & |e_1 h_3 \rangle & |e_2 h_3 \rangle
\\ |e_1 h_1 \rangle & 1 & 0.9 & 0.1 & 0.05 & 0 & 0
\\ |e_2 h_1 \rangle & 0.9 & 1 & 0.05 & 0.1  & 0 & 0 \\ |e_1 h_2 \rangle & 0.1 & 0.05 &
0.2 & 0.18  & 0 & 0 \\ |e_2 h_2 \rangle & 0.05 & 0.1 & 0.18 & 0.2 & 0 & 0 \\ |e_1
h_3 \rangle & 0 & 0 & 0 & 0 & 0 & 0 \\ |e_2 h_3 \rangle & 0 & 0 & 0 &
0 & 0 & 0
\end{array} \right) \;, \label{eqq3}
\end{equation}
where the matrix elements associated with the hole state $|h_3\rangle$ is set to be zero because it is 
the swept-out state into the gate.  The short-range electron-hole exchange interaction is small since the
spatial overlap between the $|h_3\rangle$ state and other states is negligible and the long-range 
electron-hole exchange interaction due to the dipole-dipole interaction may also be small.  The numerical
values in Eq. (\ref{eqq3}) are given semi-empirically.  

\subsection{Purity and fidelity of quantum state transfer}

 The time evolution of the whole system can be examined by the numerical integration of the equation
(\ref{eqq2}).  The initial state is supposed to be given by Eq. (\ref{eqq4}).  After a sequence of 
tunneling processes, 
the hole is eventually settled into the state $|h_3\rangle$. At this stage we are concerned with the
quantum coherence of the electron spin state. To examine the quantum coherence we calculate the reduced
density matrix of the electron by taking the trace over the hole states of the density matrix for the whole system :
\begin{equation}
\rho_{{\rm electron}}={\rm Tr}_{{\rm hole}} \; \rho \;. \label{eqq6}
\end{equation}
Then the purity of the electron state is estimated by
\begin{equation}
{\cal P}={\rm Tr} \; \rho^2_{{\rm electron}} \;.
\end{equation} 
We compared the purity for two cases of favorable and unfavorable conditions for the quantum state transfer. In the
favorable case, the magnitude of the electron-hole exchange interaction denoted by
$W$ is taken to be 3 $\mu$eV and the hole transfer (tunneling) time ($1/t_{12}=1/t_{23}=$1 ps) is short enough 
to suppress the spin state mixing by the
electron-hole exchange interaction. Results are not sensitive to the 
polarization angle $\theta$ of the excitation light in Eq. (\ref{eqq5}) and are exhibited in Fig. 3. 
The purity is high enough ($> 0.9999$) over a nanosecond 
to guarantee the secure quantum state transfer.  The decay of the purity is determined by the electron
spin decoherence/relaxation times which are supposed here to be
\begin{equation}
T_1= 1/\gamma^e_1=100 \; \mu{\rm s} \;, \quad T_2=1/\gamma^e_{12}= 1 \; \mu{\rm s} \;.
\end{equation}
\begin{figure}
\begin{center}
\includegraphics[width=8cm]{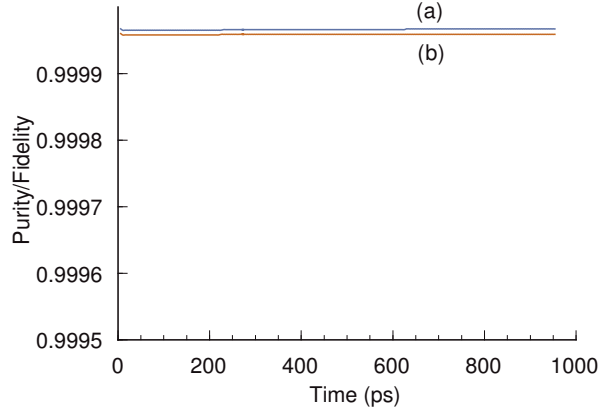}
\end{center}
\caption{{\fontsize{9pt}{9pt}\selectfont Time evolution of the purity (a) of the electron spin state and the fidelity (b) of the quantum state transfer after photoexcitation for the case of favorable conditions.}}
\end{figure} 
 In the unfavorable case, the magnitude of the electron-hole exchange interaction 
($W=20 \; \mu$ eV) is rather large and the hole tunneling time ($1/t_{12}=1/t_{23}=$10 ps) is not short enough to suppress 
the spin state mixing by the electron-hole exchange interaction. In this case also, results are not 
sensitive to the polarization angle of the excitation light and are exhibited in Fig. 4. 
The purity decreases rapidly
within several tens of picoseconds after photoexcitation due to the electron-hole exchange interaction. 
The characteristic time scale is given by $\hbar/W$ and is 33 ps. Then 
the purity is decreasing slowly due to the electron spin decoherence itself ($\sim 1 \; \mu$s). 
\begin{figure}
\begin{center}
\includegraphics[width=8cm]{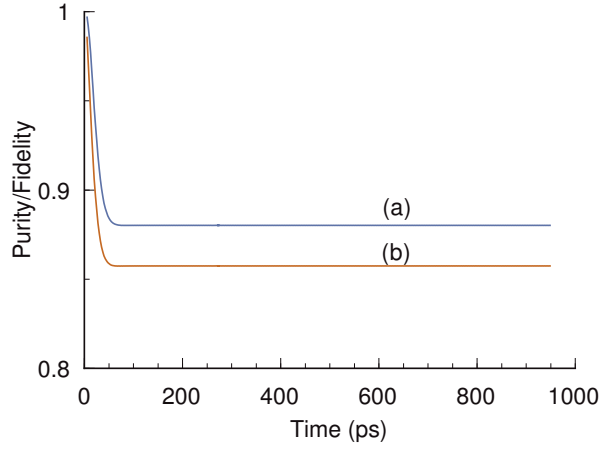}
\end{center}
\caption{{\fontsize{9pt}{9pt}\selectfont Time evolution of the purity (a) of the electron spin state and the fidelity (b) of the quantum state 
transfer after photoexcitation for the case of unfavorable conditions.}}
\end{figure}

 We also examined the fidelity of the quantum state transfer. Ideally we want to prepare the electron spin state
corresponding to the photon polarization state in Eq. (\ref{eqq5}) as
\begin{equation}
\rho_{{\rm electron}}^0=\cos^2 \theta |e_1\rangle\langle e_1| + \sin^2 \theta |e_2\rangle\langle e_2|
+i \sin \theta \cos \theta |e_2\rangle\langle e_1|-i\sin \theta \cos \theta |e_1\rangle\langle e_2| \;.
\end{equation}
The fidelity is defined by 
\begin{equation}
{\cal F}(t)={\rm Tr} \; \rho_{{\rm electron}}^0 \; \rho_{{\rm electron}}(t) \;,
\end{equation}
where $\rho_{{\rm electron}}(t)$ is the reduced density matrix of the electron defined in Eq. (\ref{eqq6}).
This quantity indicates to what extent the actual state is close to the ideally prepared state. 
The results are exhibited by curves (b) in Figs. 3 and 4. Qualitative features are the same as in the case of purity.

 Based on these results, we can conclude that for the secure quantum state transfer we have to extract the 
photoexcited hole quickly before the spin state mixing sets in due to the electron-hole exchange interaction.
The model used here has an empirical character and a more quantitative analysis would be necessary for the definite design of experiments of quantum state transfer\cite{kos3,rik}.

\section{QUANTUM CORRELATION (Bell) MEASUREMENT BETWEEN TWO ELECTRONS}

 In the scheme of quantum repeater, the primary elements are the quantum state transfer between a photon 
and an electron and the entanglement swapping through the Bell (correlation) measurement between two 
electrons which are created through the quantum state transfer from two photons. It is preferable to 
do the Bell measurement between electrons instead of photons because the mismatch between the photon 
arrival times can be compensated by the rather long coherence time of electrons, whereas the storage 
of photons is rather difficult although the techniques for the photon storage are progressing steadily. 
Thus we start the discussion assuming that two electrons are prepared in a semiconductor nanostructure, 
e.g., a quantum dot. We propose an optical method to measure the spin state of two electrons based on 
the Faraday or Kerr rotation. Here we employ a linearly polarized off-resonant probe light and measure 
the orientation of the transmitted (reflected) light. Thus the method can be non-destructive in the same 
sense as demonstrated for the case of a single electron\cite{ber,ata1}.

\begin{figure}[!t]
\begin{center}
\includegraphics[width=12cm,clip]{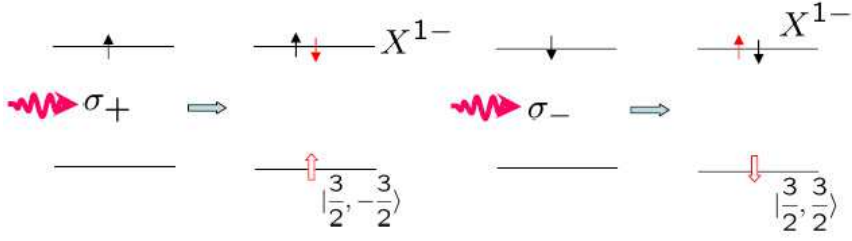}
\end{center}
\caption{{\fontsize{9pt}{9pt}\selectfont Elementary processes of the Faraday rotation for the case of a single resident electron. 
$\sigma_{+(-)}$ denotes the right (left) circularly polarized light. The upper (lower) horizontal line 
indicates the electron (hole) level. A thin (thick empty) arrow represents an electron (a hole) with 
the spin direction along the arrow. }} \label{fara1}
\end{figure}

 Before going into details, let us review briefly the elementary processes of the Faraday rotation 
for the case of a single electron. We consider a III-V semiconductor quantum dot in which the hole 
ground state is the heavy hole state and a magnetic field is applied along the crystal growth
 direction (namely, perpendicular to the quantum well plane). As is well known, the right-circularly polarized light denoted by $\sigma_+$ excites a down spin electron from the valence band state $|3/2, -3/2\rangle$ creating a charged exciton or trion, while the left-circularly polarized light denoted by $\sigma_-$ excites an up spin electron from the valence band state $|3/2, 3/2\rangle$, as exhibited in Fig. \ref{fara1}.  When we probe the system with a linearly polarized light along the $x$ direction, i.e.,
\begin{equation}
|x\rangle=\frac{1}{\sqrt{2}}(|\sigma_+\rangle+|\sigma_-\rangle) \;,
\end{equation}
one of the circular components receives a phase shift and the Faraday rotation occurs. Thus we can distinguish the two spin states of an electron by the sign of the Faraday rotation angle. 
 
\begin{figure}[!t]
\begin{center}
\includegraphics[width=12cm,clip]{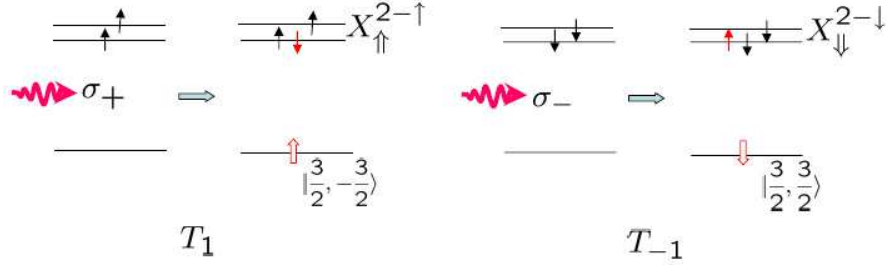}
\end{center}
\caption{{\fontsize{9pt}{9pt}\selectfont Elementary processes of the Faraday rotation for the triplet $T_1$ and $T_{-1}$ states of two resident electrons.}} \label{fara2}
\end{figure}

\begin{figure}[!t]
\begin{center}
\includegraphics[width=12cm,clip]{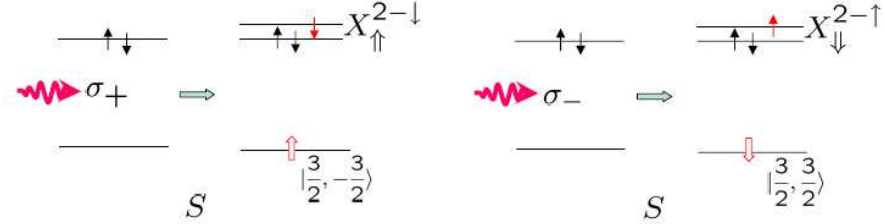}
\end{center}
\caption{{\fontsize{9pt}{9pt}\selectfont Elementary processes of the Faraday rotation for the singlet $S$ state of two resident electrons.}}
\label{fara3}
\end{figure}

\begin{figure}[!t]
\begin{center}
\includegraphics[width=12cm,clip]{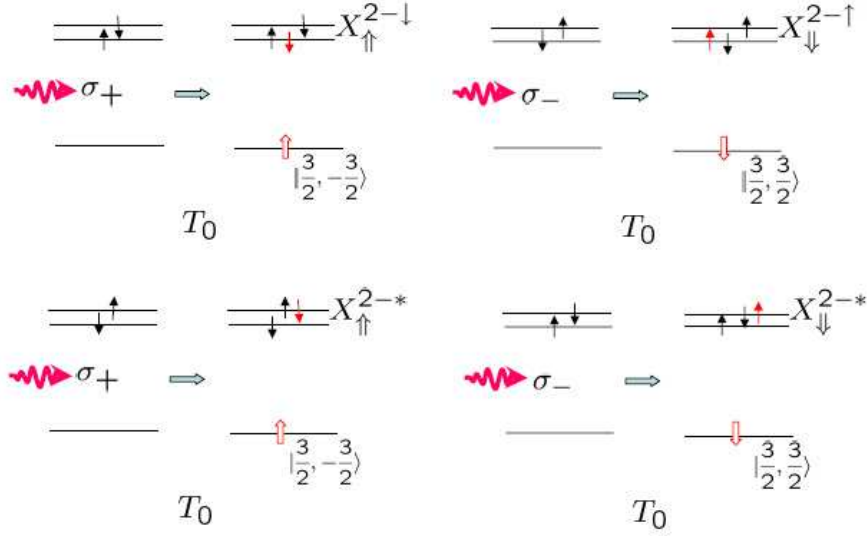}
\end{center}
\caption{{\fontsize{9pt}{9pt}\selectfont Elementary processes of the Faraday rotation for the triplet $T_0$ state of two resident electrons. $X_{\Uparrow}^{2-*}$ and $X_{\Downarrow}^{2-*}$ denote excited states of the doubly negatively charged exciton. }} \label{fara4}
\end{figure}

  Now we extend this argument to the case of two electrons and consider relevant elementary processes for four states of two electrons, namely, the singlet state(S) and the triplet states with the magnetic quantum number 1, 0 and -1 ($T_1, T_0, T_{-1}$). For the $T_1$ state, spins of the two resident electrons are aligned in the same direction and a $\sigma_+$ polarized light excites a down spin electron from
the valence band creating a doubly negatively charged exciton $X^{2-}$, as shown in Fig. \ref{fara2}, in which the lowest electron orbital state is occupied by a spin-singlet electron pair and the spin direction of the electron in the second lowest orbital state is indicated in the superscript and the spin direction of the hole is depicted in the subscript. This $T_1$ state is optically inactive for the $\sigma_-$ polarized light. For the $T_{-1}$ state, a $\sigma_-$ polarized light excites an up spin electron from the valence band creating another doubly negatively charged exciton. This $T_{-1}$ state is optically inactive for the $\sigma_+$ polarized light. Thus these two states can be distinguished by the sign of the Faraday rotation angle. On the other hand, the S and $T_0$ states are optically active for both circular polarizations as exhibited in Figs. \ref{fara3} and \ref{fara4}, and the sign of the Faraday rotation angle is determined by the competition between the phase shifts for each circular component.

The expression of the Faraday rotation angle is obtained in the perturbation theory and is composed of two terms: 
\begin{equation}
\varphi \propto \sum_{j(\sigma +)} \frac{|\langle j|P_{\sigma+}|i\rangle|^2 
(E_{j, i}-\hbar \omega)}{(\hbar^2 \gamma^2_{j,i}+(E_{j, i}-\hbar \omega)^2)}
 - \sum_{k(\sigma -)} \frac{|\langle k|P_{\sigma-}|i\rangle|^2 
(E_{k, i}-\hbar \omega)}{(\hbar^2 \gamma^2_{k,i}+(E_{k, i}-\hbar \omega)^2)} \;, \label{faraeq}
\end{equation}
where $i$ indicates the initial state of two electrons, $j\;(k)$ the final state of the optical transition
for the 
$\sigma_+\;(\sigma_-)$ component, $E_{a, b}=E_a-E_b$ with $E_a$ being the energy of the $a$ state, $\gamma_{a, b}$ the dephasing rate corresponding to the $a \leftrightarrow b$ transition and $\hbar \omega$ denotes the photon energy of the linearly polarized probe light.
As mentioned before, for the $T_1$ state only the $\sigma_+$ transitions  
contribute, whereas for the $T_{-1}$ state only the $\sigma_-$ transitions contribute. Thus the two 
states can be distinguished by the sign of the Faraday rotation angle.  For the S and $T_0$ states, 
both $\sigma_+$ and $\sigma_-$ transitions contribute and thus more detailed arguments are necessary 
to determine the sign of the Faraday rotation angle. Now we examine the resonance position of the 
Faraday rotation angle with respect to the probe photon energy $\hbar \omega$. From the elementary 
processes exhibited in Figs. \ref{fara2}-\ref{fara4}, it is seen that for the triplet states the resonance occurs at around the energy of the doubly 
charged exciton states ($E(X^{2-})$). On the other hand, for the singlet state the resonance occurs at a higher energy than $E(X^{2-})$ because the lowest orbital state is already occupied by 
a spin-singlet electron pair and the optical transition should occur to the higher orbital state.
  
\begin{figure}[!t]
\begin{center}
\includegraphics[width=12cm,clip]{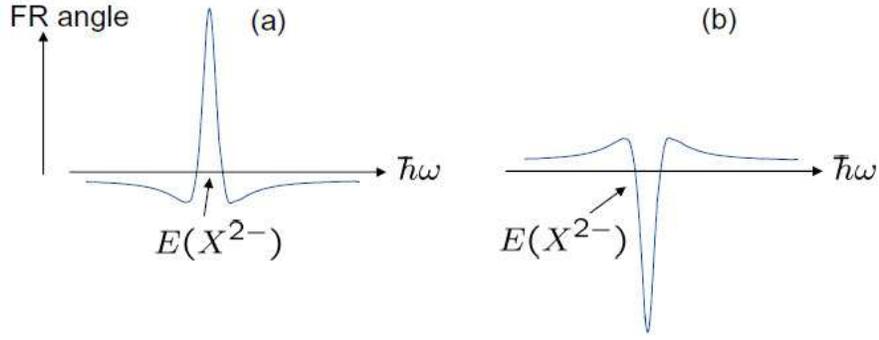}
\end{center}
\caption{{\fontsize{9pt}{9pt}\selectfont Dependence on the probe photon energy ($\hbar\omega$) of the Faraday rotation angle for the triplet $T_0$ state and the singlet $S$ state of two resident electrons. It depends on the sign of the energy difference; namely, (a) $|E(X^{2-\downarrow}_{\Uparrow})-E(X^{2-\uparrow}_{\Downarrow})| > 0$, (b) $|E(X^{2-\downarrow}_{\Uparrow})-E(X^{2-\uparrow}_{\Downarrow})| < 0$ }} \label{fara5}
\end{figure}

 Now we discuss more details of the Faraday rotation angle for the case of $T_0$ state. As mentioned before, both $\sigma_+$ and $\sigma_-$ circular components contribute to the Faraday rotation. The lowest-energy final state of the optical transition for each circular component is given by
\begin{equation}
j(\sigma+)=X^{2-\downarrow}_{\Uparrow} \;, \qquad k(\sigma-)=X^{2-\uparrow}_{\Downarrow} \;.
\end{equation}
The energies of these states are different in a magnetic field because the spin configuration is different for
these states. In terms of the electron $g$-factor $g_{c(v)}$ for the conduction (valence) band, these energies are given as
\begin{eqnarray}
&&E(X^{2-\downarrow}_{\Uparrow})\cong -\frac{1}{2}(g_c \mu_B B - g_v \mu_B B) 
+E_0 \;, \\ 
&& E(X^{2-\uparrow}_{\Downarrow})\cong  \frac{1}{2}(g_c \mu_B B - g_v \mu_B B) 
+E_0 \;,
\end{eqnarray}
where $E_0$ is the lowest energy of the interband transition. Then the energy difference 
$|E(X^{2-\downarrow}_{\Uparrow}) -E(X^{2-\uparrow}_{\Downarrow})|$
is typically about one tenth of meV for a magnetic field about 1 Tesla and is comparable to the 
dephasing rate of the optical transitions. From the formula in Eq. (\ref{faraeq}) we see that the dependence of the Faraday rotation angle on the probe photon energy
is determined by the difference between two dispersive curves with nearly equal resonance energies. Thus
the profile is given by the derivative of the dispersive curve as shown in Fig. \ref{fara5}, depending on the sign of the energy difference. The same situation holds for the singlet state $S$. 

\begin{figure}[!t]
\begin{center}
\includegraphics[width=12cm,clip]{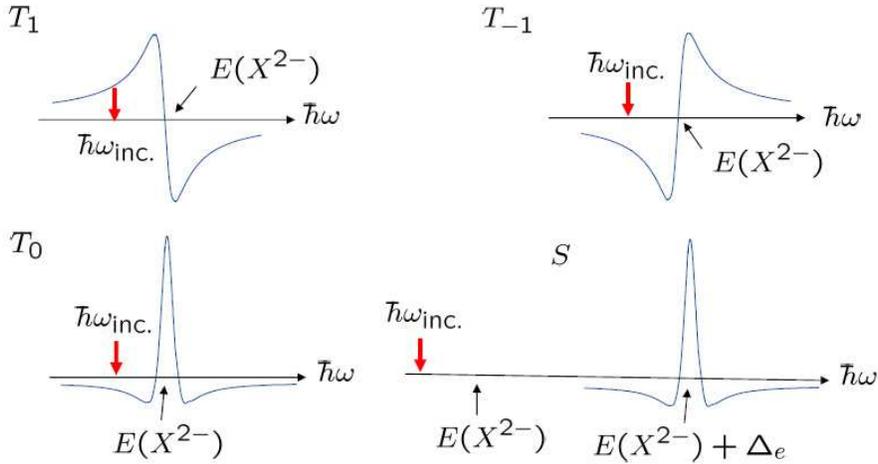}
\end{center}
\caption{{\fontsize{9pt}{9pt}\selectfont Dependence on the probe photon energy ($\hbar\omega$) of the Faraday rotation angle for the three triplet states $T_1, T_0, T_{-1}$ and the singlet state $S$ of two resident electrons. Those for $T_0$ and $S$ are exhibited for the case of $|E(X^{2-\downarrow}_{\Uparrow})-E(X^{2-\uparrow}_{\Downarrow})| > 0$.}}
\label{fara6}
\end{figure}

Summarizing these considerations, we can show the schematic dependence of the Faraday rotation angle 
on the probe photon energy in Fig. \ref{fara6}. The triplet states $T_1$ and $T_{-1}$ exhibit a typical 
dispersive lineshape. On the other hand, the profile for the triplet $T_0$ and the singlet $S$ states 
is given by the derivative of the dispersive curve, where the case of 
$|E(X^{2-\downarrow}_{\Uparrow})-E(X^{2-\uparrow}_{\Downarrow})| > 0$ is assumed. The resonance occurs 
at around the energy of the doubly charged exciton state denoted by $E(X^{2-})$ for the triplet states, 
whereas for the singlet state it occurs at a higher energy than $E(X^{2-})$ by the orbital excitation energy $\Delta_e$. 
Thus when we choose the probe photon energy at the downward arrow as shown in Fig. \ref{fara6}, the Faraday rotation angle is positive for the $T_1$ state and is negative for the $T_{-1}$ state. For the $T_0$ state, the Faraday rotation angle is negative but the magnitude is small. For the singlet $S$ state, the Faraday rotation angle would be vanishingly small because of the large off-resonance. Consequently, we can distinguish between the four states of two electrons by the magnitude and the sign of the Faraday rotation angle. 

  Now we discuss relevant parameters to optimize the Faraday rotation measurement. The essential requirement is the preparation of the lowest two orbital states which are energetically well-separated from higher excited states. We consider a circularly symmetric GaAs quantum dot with parabolic lateral confinement under a magnetic field along the growth direction. Then the orbital eigenstates are represented by the Fock-Darwin states\cite{din1,din2} whose eigenenergies are given by
\begin{eqnarray}
&&E_{\nu, n}=(|n|+1+ 2\nu)\hbar \Omega + \frac{n}{2} \hbar \omega_c \\
{\rm with} \;\; && \Omega=\sqrt{\omega^2_0 +
\frac{\omega^2_c}{4}} \;, \; \omega_c=\frac{eB}{m^* c} \;,
\end{eqnarray}
where $\omega_0$ is the frequency of the harmonic confinement in the lateral direction and $m^*$ is the electron effective mass.
When we employ the parameter values: $\hbar\omega_0=5$meV, $B=5$T and $m^*=0.067 m_0$ with $m_0$ being the free electron mass, we have $\hbar\omega_c=8.7$meV and $\hbar\Omega=6.63$meV. The lowest two orbital levels have the spacing of 2.3 meV and are well-separated from the higher orbital level by 8.7 meV. These parameter values would enable the Faraday rotation measurement to be carried out reliably.

\section{DYNAMICS IN ELECTRON-NUCLEI COUPLED SYSTEM}

 As mentioned in Sec. 1, the electron spin decoherence time reported so far for low temperatures 
is about a few microsecond and is not sufficiently long for the secure quantum information processing. 
Here the hyperfine (HF) interaction with the host nuclei\cite{abr,mer} is considered to
be the main decoherence mechanism, dominating over
spin-orbit interactions which act on a time scale of tens of milliseconds\cite{kro,meu} or 
even longer. Consequently there have been proposals to reduce the HF induced decoherence by measuring or 
polarizing the nuclear
spins\cite{tay,ima1,kla,gie,ste} and to use nuclear spins as 
a quantum memory\cite{tay1}. 
 
 Extending these studies, we investigate the electron-nuclei spin coupling in quantum dots (QDs) and show that 
consecutive measurements of the electron spin state 
 following the HF interaction are correlated and lead to purification of the nuclear spin system.
More specifically, starting from an unknown initial state of nuclear
spins, successive measurements of the electron spin state result in narrowing of the
distribution of the nuclear spin field.
We predict that the purification of the nuclear spin state
 would lead to the bunching of results of the electron spin state
measurements and also to the reduction in the electron
spin decoherence induced by the HF interaction. 
For the physical realization of the proposals we will in particular discuss a double QD occupied by two 
electrons, and a single QD occupied by one or two electrons.
Under sufficiently high magnetic fields compared with the effective HF field (Overhauser field), these 
systems provide the desired two-level system with a unidirectional HF field.

First of all we consider an electrically gated double QD occupied by two electrons\cite{pet,coi}.
The excited electronic orbitals of QDs have an energy much greater than the thermal energy and the adiabatic 
voltage sweeping rates, so that the electrons occupy only the ground state orbitals.
Under a high magnetic field where the electron Zeeman splitting is much greater than
 the HF fields and the exchange energy, dynamics takes place in the spin singlet ground 
state $|S\rangle$ 
and triplet state of zero magnetic quantum number $|T\rangle$. For the singlet state each electron can be 
found in the different QD or both in the same QD, whereas for the triplet state electrons can only be found 
in different QDs.
In order to derive the effective Hamiltonian for the singlet and triplet states of two electrons each of which is lying in a different QD, we rewrite the HF interaction for two electrons in the bases of singlet and triplet states and find that the mean HF field induces mixing within triplet states and the difference of the HF fields in two QDs induces coupling between the singlet and triplet states. When the Zeeman energy is much larger than the HF fields, the coupling terms among the triplet states and those between $|T_{\pm}\rangle$ (triplet states with the magnetic quantum number of 1 or -1) and the singlet state $|S\rangle$ can be neglected and the HF interaction reduces to
\begin{equation}
V_{HF}=\delta h_z(|S\rangle\langle T|+{\tt h.c.})/2  \label{hfz}
\end{equation}
with $\delta h_z=h_{Lz}-h_{Rz}$ being the difference of the HF fields along the applied field direction in the left and right QDs. Including the exchange energy splitting $J$ between $|S\rangle$ and $|T\rangle$, the effective Hamiltonian is given as
\begin{eqnarray}
H_e=J S_z+ r \delta h_z S_x \;, \label{eq_hf}
\end{eqnarray}
where ${\bf S}$ is the pseudospin operator with $|T\rangle$ and $|S\rangle$ forming the $S_z$ basis and the parameter $r$ characterizes the hybridization ratio of the singlet state whose two electrons are separated in different QDs in the true singlet ground state.
When both electrons are localized in the same dot, 
$r\rightarrow 0$ and $J\gg \delta h_z$. On the other hand, when they are located in different dots, the HF coupling is maximized $r\rightarrow 1$ and $J\rightarrow 0$.

 In the following we develop the arguments based on Eq. (\ref{eq_hf}) assuming that the time scale of the nuclear spin variation is much longer than that of the electron spin measurements. Thus we take into account only the static(inhomogeneous) distribution of the nuclear Overhauser field but not the dynamical motion of the nuclear spin system itself due to the dipole-dipole interaction which would induce additional decoherence of the electron spins.

\subsection{Bunching in electron spin measurements}

 Now we show that by electron spin measurements in a double QD governed by Eq. (\ref{eq_hf}), the coherent 
behavior of nuclear spins can be demonstrated.
 Electron spins are initialized in the singlet state and the nuclear spin states are initially in a mixture 
of  $\delta h_z$ eigenstates:
\begin{equation}
\rho(t=0)=\sum_n p_n \;\rho_n \;|S\rangle\langle S| \;,
\end{equation}
where $\rho_n$ is a nuclear eigenstate with an eigenvalue $h_n$:
\begin{equation}
 {\rm Tr} \; \rho_n \;\delta h_z =h_n \:, \quad {\rm Tr} \; \rho_n=1 \;.
\end{equation} 
$p_n$ is the probability of the hyperfine field $\delta h_z$ having the value $h_n$.
In the unbiased regime $r=1$, the nuclear spins and the electron spins interact for a time span of $t$.
The time evolution of the system is described as follows:
\begin{eqnarray}
&&\rho(t=0)=\sum_n p_n \rho_n \; |S\rangle\langle S| \rightarrow 
\rho(t)=\sum_n p_n \rho_n \; |\Psi_n\rangle\langle \Psi_n| \;, \\
&& |\Psi_n\rangle=\alpha_n(t) |S\rangle + \beta_n(t) |T\rangle \;, \\
&& \alpha_n(t)=\cos\Omega_n t/2+iJ/\Omega_n \sin\Omega_n t/ 2 \;, \;  \beta_n(t)=
 -ih_n/\Omega_n\sin\Omega_n t/2\;, \label{eq41} \\
&& \Omega_n=\sqrt{J^2+h_n^2} \;.
\end{eqnarray}
Then the gate voltage is swept adiabatically, switching off the HF interaction $r\rightarrow 0$, in a time scale much shorter than HF interaction time.
Next a charge state measurement is performed which detects a singlet or triplet state. Probability to 
detect the singlet state is 
\begin{equation}
P_S(t)={\rm Tr}_{{\rm nuc.}} \langle S|\rho(t)|S \rangle=\sum_n p_n|\alpha_n|^2 \;,
\end{equation}
where the nuclear states are traced out because they are not observed. In the same way
the probability to detect the triplet state is 
\begin{equation}
P_T(t)=\sum_n p_n|\beta_n|^2 \;.
\end{equation}
After the first measurement, assuming that the outcome is the singlet, the system is in the state given by
\begin{equation}
\frac{1}{P_S(t)} \sum_n p_n |\alpha_n|^2 \rho_n |S \rangle \langle S| \;,
\end{equation}
where $P_S(t)$ in the denominator is the normalization constant of the density matrix. In the second run we again initialize the system in the spin singlet state of the two electrons, preparing the state:
\begin{equation}
\rho'(t=0)=\frac{1}{P_S(t)} \sum_n p_n |\alpha_n|^2 \rho_n \; |S\rangle \langle S|
\end{equation}
and turn on the HF interaction. After the same period $t$ as in the first run, the system evolves to
\begin{eqnarray}
&&\rho'(t)=\frac{1}{P_S(t)} \sum_n p_n |\alpha_n|^2 \rho_n \;|\Psi_n\rangle \langle \Psi_n| \label{eq42} \\
&&|\Psi_n\rangle=\alpha_n(t) |S\rangle + \beta_n(t) |T\rangle \;,
\end{eqnarray}
where $\alpha_n(t)$ and $\beta_n(t)$ are given in Eq. (\ref{eq41}).
Then the probability to detect the singlet state in the second measurement $P_{SS}(t; t)$, where the first (second) subscript represents the outcome of the first (second) measurement, is given by
\begin{equation}
P_{SS}(t; t)= P_S(t) \cdot {\rm Tr}_{{\rm nuc.}} \langle S|\rho'(t)|S \rangle = \sum_n p_n |\alpha_n|^4 \;,
\end{equation}
where it is to be noted that the probability $P_S(t)$ of the first measurement cancels the normalization factor in Eq. (\ref{eq42}). In the same way the probability to detect the triplet state in the second measurement $P_{ST}(t; t)$ is given by
\begin{equation}
P_{ST}(t; t)= \sum_n p_n |\alpha_n|^2 |\beta_n|^2 \;.
\end{equation}
After the second measurement, assuming that the first (second) outcome is the singlet (triplet), the system is in the state given by
\begin{equation}
\frac{1}{P_{ST}(t;t)} \sum_n p_n |\alpha_n|^2 |\beta_n|^2 \rho_n 
\; |T\rangle \langle T| \;,
\end{equation}
where $P_{ST}(t; t)$ in the denominator is the normalization factor of the density matrix. In this way we can repeat many times of measurements.

In general over $N$ measurements, the nuclear state conditioned on $k\;(\leq N)$ times singlet and $N-k$ 
times triplet detection is 
\begin{equation}
\sigma_{N,k}=\bigl(^N_{\,k}\bigr)\sum_n p_n|\alpha_n|^{2k}|\beta_n|^{2(N-k)}\rho_n\;, \label{nucstate}
\end{equation}
the trace of which yields the probability of $k$ times singlet outcomes
\begin{align}
P_{N,k}={\rm Tr} \; \sigma_{N,k}
 =\bigl(^N_{\,k}\bigr)\langle|\alpha|^{2k}|\beta|^{2(N-k)}\rangle \;, \label{Pqm}
\end{align}
where  $\langle\ldots\rangle$ is the ensemble averaging over the hyperfine field $h_n$\cite{mer}.
 Hereafter, this case will be referred to as the {\it coherent regime}.
One can contrast this regime with the {\it incoherent regime} in which nuclear spins lose 
their coherence between the successive spin measurements and relax to the equilibrium distribution. The result for the latter regime is given by
\begin{equation}
 P'_{N,k}=\bigl(^N_{\,k}\bigr)\langle|\alpha|^2\rangle^{k}\langle|\beta|^2\rangle^{(N-k)} \;. \label{Psc}
\end{equation} 

In the following we assume that the initial nuclear spins are unpolarized and randomly oriented and thus the distribution of the hyperfine field is characterized by a Gaussian distribution with variance $\sigma^2$:
\begin{equation}
p[h]=\frac{1}{\sqrt{2\pi\sigma^2}}e^{-\frac{h^2}{2\sigma^2}}\;.
\end{equation}
The summation is converted to an integration: 
\begin{equation}
\sum_n \; p_n \ldots \rightarrow \int {\tt d}h \, p[h] \ldots \;.
\end{equation}
As the simplest case, let us examine the results for two ($N=2$) measurements, each following a HF interaction of 
duration time $t$. The probability for two consecutive singlet detections is given by
\begin{equation}
P_{2,2}=\langle|\alpha|^4\rangle=\{6+2 e^{-2\bar{t}^2}+8e^{-\bar{t}^2/2}\}/16 
\end{equation}
for the coherent regime with $\bar{t}=\sigma t$ and this is always greater than
\begin{equation} 
P'_{2,2}=\langle|\alpha|^2\rangle^2=\{4+8e^{- \bar{t}^2/2}+4e^{-\bar{t}^2}\}/16 
\end{equation}
for the incoherent regime. These results are given particularly for $J=0$.
As $J$ is increased the probabilities approach each other and for $J\gg \sigma$ they become 
identical\cite{cak}.
\begin{figure}[!t]
\begin{center}
\includegraphics[width=12cm]{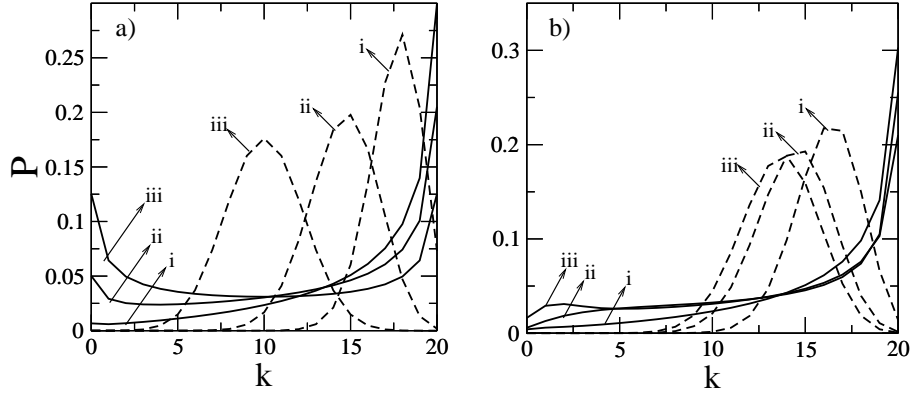}
\end{center}
\caption{{\fontsize{9pt}{9pt}\selectfont Probability distribution for $N=20$ measurements as a function of times ($k$) of singlet 
detections, for coherent regime (solid lines) and incoherent regime (dashed lines). Two cases of the 
exchange energy are considered  a) $J=0$ , and b) $J/\sigma=0.5$ , for HF interaction times 
$\sigma\tau=$ i) $0.5$, ii) $1.5$, and iii) $\infty$. }}  \label{Fig_20meas}
\end{figure}
In Fig. \ref{Fig_20meas}, for $N=20$ measurements, $P_{N,k}$ and $P'_{N,k}$ are shown
 for HF interaction times $\sigma\tau=0.5, 1.5, {\rm and}\;  \infty$. For $\tau=0$, both probabilities  
 are peaked at
$k=20$. However, immediately after the HF interaction is introduced, the probability distributions show 
distinct behavior. The measurement results in the incoherent regime approach a Gaussian distribution. 
In the coherent case the probabilities
bunch at $k$=0 and 20 for $J=0$, and when $J/\sigma=0.5$ those bunch at $k=20$ only. As $J$ is increased 
above some critical value, no bunching takes place at $k=0$ singlet measurement. When $J$ is finite, the singlet state is energetically favored and the bunching of the singlet state outcome is more probable. Thus  
if the nuclear spins are coherent over the span of the experiment,
successive electron spin measurements are likely to be biased to all singlet or triplet outcomes.

We discuss in brief the feasibility to observe the
predicted phenomenon. The duration of the cycle involving electron
spin initialization and measurement is about $10~
\mu$s\cite{pet}. The nuclear spin coherence time
determined mostly by the nuclear spin diffusion is longer than about
several tens of ms\cite{pag}. For a HF interaction of duration $\tau=4\sigma^{-1}\sim 40\mu$s 
\cite{pet} in each step, the bunching can be observed for $N$ successive
measurements up to $N\sim 200$.

\subsection{Purification of nuclear spin state and revival of the initial electron state}

\begin{figure}[!t]
\begin{center}
\includegraphics[width=12cm]{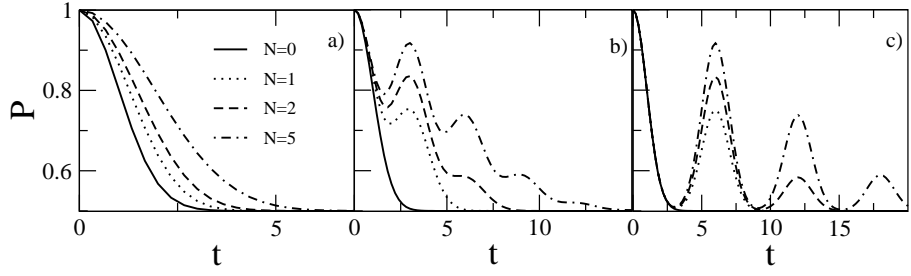}
\end{center}
\caption{{\fontsize{9pt}{9pt}\selectfont Conditional probability for singlet state detection as a function of HF interaction time 
$\sigma t$, subject to $N=0,1,2,5,10$ times prior singlet state measurements and for HF interaction 
times of a) $\sigma\tau=1.0\;$, b) $\sigma\tau=3.0\;$, and c) $\sigma\tau=6.0\;$. }} \label{Fig-cond}
\end{figure}

The nuclear spin state conditioned on the previous electron spin measurements is no longer random even if 
they are initially random. To be stated in more detail, the quantum correlation is built up in the nuclear spin system as a consequence of back-actions of the electron spin measurements. This quantum correlation affects in turn the outcomes of the 
electron spin measurements. In order to examine this situation, we 
consider the case: Starting from a random spin configuration, $N$ successive
electron spin measurements are performed, each following initialization of electron spins in the spin 
singlet state and a HF interaction of duration $\tau_i$ ($i=1\ldots N$) and all the outcomes turn out 
to be the 
singlet. Then again the HF interaction is switched 
on for a time $t$, and the $(N+1)$-th measurement is carried out. The conditional probability to recover 
the initial state, namely to observe again the singlet state, is given by
\begin{eqnarray}
 P(t)=\frac{\sum\prod_{i=1}^{N+1}\binom{2}{s_i}e^{-\frac{1}{2}\bigl[\sum_{j=1}^N(s_j-1)\bar{\tau}_j+
(s_{N+1}-1)\bar{t}\bigr]^2}}{
4\sum \prod_{i=1}^{N}\binom{2}{s_i}e^{-\frac{1}{2}\bigl[\sum_{j=1}^N(s_j-1)\bar{\tau}_j\bigr]^2} }\;,
 \label{condprob}
\end{eqnarray}
where the sums run over $s_i=0,1,2$ and $\bar{\tau}_i=\sigma\tau_i$.
For the particular case of $\tau_1=\tau_2=\ldots=\tau_N=\tau \gg1/\sigma$, Eq. (\ref{condprob}) can be approximated as
\begin{equation}
P(t) \simeq 1/2+\sum_{s=0}^{N}(
^{2N}_{~s})e^{\frac{-\sigma^2}{2}(t-(N-s)\tau)^2}/2(^{2N}_{~N}) \quad {\rm for} \; \tau\gg 1/\sigma \:.
\end{equation}
This indicates a periodical recurrence of the initial state of the electron spin.
Indeed the initial state is revived at $t=n\tau, \,(n=1,2,\ldots,N)$ with a decreasing amplitude:
\begin{equation}
1/2+\binom{2N}{N-n}/2 \binom{2N}{N}\;.
\end{equation}
In Fig. \ref{Fig-cond} the conditional probabilities in Eq. (\ref{condprob}) are shown for $\sigma\tau=1.0, 3.0, 
6.0$ subject to $N=0,1,2,5,10$ times prior singlet measurements in each. Revivals or recurrences are observable only for 
$\sigma\tau>1$, because the modulation period of the nuclear field spectrum characterized by $1/\tau$ 
should be smaller than the variance $\sigma$, as will be explained later. 

In order to understand the mechanism of revivals, we examine the purity of the nuclear spin system.
 The purity of a system
characterized by the density matrix $\rho$ is given by 
\begin{equation}
{\cal P}={\rm Tr} \; \rho^2 \;.
\end{equation} 
We consider the nuclear spin state prepared by $N$ successive electron spin measurements
in which all the outcomes are the singlet and the duration times of the HF interaction are  
$\tau_{1}, \cdots,\tau_N$. The purity of this nuclear spin state is calculated as
\begin{eqnarray}
{\cal P}=\frac{1}{\cal D}\frac{\sum_{s_i=0}^
4\prod_{i=1}^{N}\binom{4}{s_i}e^{-\frac{1}{2}\bigl[\sum_{j=1}^N(s_j-2)\bar{\tau}_j\bigr]^2}}
{\bigl[ \sum_{s_i=0}^
2 \prod_{i=1}^{N}\binom{2}{s_i}e^{-\frac{1}{2}\bigl[\sum_{j=1}^N(s_j-1)\bar{\tau}_j\bigr]^2}\bigr]^2 } \;,
\label{purity}
\end{eqnarray}
where ${\cal D}$ is the dimension of the Hilbert space for the nuclear spins. For a fixed ratio of 
$\tau_1:\tau_2:\ldots:\tau_N$, the purity in 
Eq. (\ref{purity}) is a monotonically increasing function of time. For
$\bar{\tau}_i=\sigma\tau_i \gg 1$, one can attain various asymptotic limits for
the purity.  For instance, for $N=2$, there are three asymptotic limits:

\noindent a) $\tau_1=2\tau_2$  
\begin{equation}
{\cal P}=11/4{\cal D}\;,
\end{equation}
b) $\tau_1=\tau_2$  
\begin{equation}
{\cal P}=35/18{\cal D}\;,
\end{equation}
c) otherwise 
\begin{equation}
{\cal P}=9/4{\cal D}\;.
\end{equation}
For $N=2$ with $\tau_1=2\tau_2=2\tau\gg 1/\sigma$,  the conditional
probability in Eq. (\ref{condprob}) is given as
\begin{equation}
P(t) \simeq 1/2+ \sum_{n=0}^3 (4-n)\exp[-(\bar{t}-n\bar{\tau})^2/2 ]/8 \;,
\end{equation}
whereas for $\tau_1=\tau_2=\tau\gg 1/\sigma$,
\begin{equation}
P(t) \simeq 1/2+\bigl\{ e^{-\frac{(\bar{t}-2\bar{\tau})^2}{2} }+
4e^{-\frac{(\bar{t}-\bar{\tau})^2}{2} }+6e^{-\frac{\bar{t}^2}{2} }
\bigr\}/12\;.
\end{equation}
It can be seen that as the purity of nuclear spins increases, more revivals are present
with an increased amplitude. Thus we can conclude that the mechanism of the revival phenomenon is the purification of the nuclear spin system through the electron state measurements.

Previously, various methods have been proposed in order to control the nuclear spin system. A typical method is based on measurement of the HF field in a QD with some 
precision\cite{kla,gie,ste}. In the method proposed here, on the other hand, the nuclear spin state can 
be conditionally purified without determining the precise value of the HF field. Although the HF 
field is still assuming various values, the quantum correlation is built up, leading to the revival phenomenon.
In order to understand these features clearly, we simplify the situation by putting $J=0$ in Eq. (\ref{eq41}). Then the nuclear state density matrix, after $N$ times electron spin measurements with all the outcomes being the singlet, is given by
\begin{equation}
\rho_{nuc.} \propto \sum_n p_n |\cos h_n \tau/2|^{2N} \rho_n
\end{equation}
and the corresponding nuclear field spectrum or distribution function assumes a form of
\begin{equation}
p[h] ={\cal N} e^{-h^2/2\sigma^2} \cos^{2N}[h \tau/2] \;, \label{eq71}
\end{equation}
where ${\cal N}$ is an appropriate normalization constant. Thus the modulation period of the nuclear field
spectrum is characterized by $1/\tau$. When this modulation period is larger than the width $\sigma$ of the initial
Gaussian distribution, i.e., $\sigma \tau < 1$, the effect of modulation is not manifest. This explains the 
absence of revivals in Fig. \ref{Fig-cond} a).

Since the absolute value of a cosine function is less than unity, many times multiplication of the cosine function leads to sharpening of the distribution. Indeed, we can approximate for $N \gg 1$ as
\begin{eqnarray}
&&\cos^{2N} \theta = \sum_{ s \in \mathbb{Z}} (1 - \frac{1}{2} (\theta- s \pi)^2+\cdots)^{2N} \\
&&\hspace{1.2cm} =\sum_{ s \in \mathbb{Z}} (1 - N (\theta- s \pi)^2+\cdots) \simeq
\sum_{ s \in \mathbb{Z}} \exp[- N (\theta- s \pi)^2] \;.
\end{eqnarray}
Then Eq. (\ref{eq71}) can be cast into a series of very narrow Gaussian functions
\begin{eqnarray}
&&p[h] ={\cal N} e^{-h^2/2\sigma^2} \cos^{2N}[h \tau/2] 
 \simeq {\cal N} e^{-h^2/2\sigma^2} \sum_{s\in \mathbb{Z}} e^{-(h-h_s)^2/2\sigma_m^2} \\
&& \sigma_m^{-1}=\tau \sqrt{N/2} \;, \; h_s=2s\pi/\tau \;,
\end{eqnarray}
where it is to be noted that the width of Gaussians is inversely proportional to $\sqrt{N}$.
Thus the nuclear field distribution function is squeezed by successive electron spin measurements and this is the origin of increase in the purity of the nuclear state. Then the probability to have a singlet outcome in the $(N+1)$-th measurement after the HF interaction of period $t$ is calculated as
\begin{eqnarray}
&&P(t)=\langle \cos^2 ht/2 \rangle={\cal N} \int dh\; \cos^2(ht/2) \; e^{-h^2/2\sigma^2} 
\sum_{s\in \mathbb{Z}} e^{-(h-h_s)^2/2\sigma_m^2} \\
&&\hspace{2.9cm} =\frac{1}{2}+\frac{\cal N}{2}
 \; \sum_{s\in \mathbb{Z}}  e^{-\sigma_m^2t^2/2}e^{-h_s^2/2\sigma^2} \cos h_s t \;.
\end{eqnarray}
At $t=n\tau$, all $\cos h_s t$ terms add up constructively leading to the revival phenomenon, namely, a high probability to observe the spin singlet state periodically.

 So far we have discussed the case in which the initial state of the electron pair is the singlet. But these arguments can be extended to the case of an arbitrary initial state.  When the initial 
state of the electron pair is prepared as
\begin{eqnarray}
\Psi(t=0)=\cos\frac{\theta}{2} |S\rangle+\sin\frac{\theta}{2} e^{-i\phi} |T\rangle \;,
\end{eqnarray}
the probability to recover the initial state $\Psi(t=0)$ at time $t$ is calculated as
\begin{eqnarray}
&&F={\rm Tr}_{{\rm nuc.}} \langle \Psi(t=0)|\rho(t)|\Psi(t=0)\rangle \\
&&\hspace{0.38cm} =\sin^2\theta\cos^2\phi+(1-\sin^2\theta\cos^2\phi)P(t)
\end{eqnarray}
with $P(t)$ given by Eq. (\ref{condprob}). Thus the recurrence phenomenon can be observed for an arbitrary initial state of the electron pair.

Finally, as an example, we consider the case when the nuclear spin state is prepared by five HF interaction stages 
each of which has duration $\tau=10/\sigma$ and is followed by a singlet detection of the electron 
spin state. This conditionally prepared nuclear spin state revives the initial electron state 
 at $t=s \tau \; (s=1,2,\ldots,5)$ with the probability of $11/12,31/42,\ldots,253/504$, 
respectively. Success probability to prepare 
such a state is $\sim 1/32$. Here preparation time is $T=N(\tau+\tau_w)$ where $\tau_w$ is the time 
needed for initialization and detection of the electron spin. During the HF interaction time $\tau$, 
phonon-mediated interations which act on the time scale longer than millisecond \cite{meu} 
or any other electron 
spin decoherence mechanism except the HF interaction should not take place.
 Furthermore the preparation time $T$ should be smaller than the nuclear diffusion time which is of 
the order of 10 ms\cite{pag}. Typically $\tau_w=10 \;\mu$s, and let $\sigma^{-1}=10 \; \mu$s\cite{pet}. 
For the discussed example ($\tau=10/\sigma$, $N=5$), the time needed for preparation of the desired 
nuclear spin state is $T=550 \; \mu$s which is shorter than the nuclear spin diffusion time.

\subsection{Extension to general situations}

We have so far discussed the bunching and revival phenomena only for two electrons in a double QD system. The same predictions 
can also be  made for a single QD occupied by a single electron.
Consider a single QD occupied by a single electron\cite{han,dut,ata}, under an external magnetic field where the electron 
Zeeman energy is much greater than the HF energies. Then the system is described by the Hamiltonian: 
\begin{equation}
H\simeq g_e\mu_BB S_z + h_z S_z \;,
\end{equation}
where $S_z$ is the $z$ component of the electron spin operator, $g_e$ the electron $g$-factor, $\mu_B$ the Bohr magneton and $B$ is 
the external magnetic field applied in the $z$ direction. Spin flips are suppressed since
\begin{equation} 
g_e\mu_B B\gg \sqrt{\langle {\bf h}^2\rangle}\;.
\end{equation}
The spin eigenstates along the $x$ direction are given by
\begin{equation}
|\pm\rangle=(|\uparrow\rangle\pm|\downarrow\rangle)/\sqrt{2} \;,
\end{equation}
where $|\uparrow\rangle$ and $|\downarrow\rangle$ are the eigenstates of $S_z$ and they are 
coupled via the HF interaction. Thus the relevant Hamiltonian in the basis of $|\pm\rangle$ is
\begin{equation}
H=\left(\begin{array}{cc}
|+\rangle & |-\rangle \\
0 & (g_e \mu_B B_z +h_z)/2 \\
(g_e \mu_B B_z +h_z)/2 & 0 \end{array} \right) \;.
\end{equation}
Then the spin state measurement is carried out as follows.
Each time the electron is prepared
in the $|+\rangle$ state. Next it is loaded onto the QD, then removed from the
QD after some dwelling time $\tau$. The spin measurement is
performed in the $|\pm\rangle$ basis.
 Essentially the same predictions can be made for this system
as those for the case of double QD, namely the electron
spin bunching and revival. 

We can extend the arguments also to the case of a pair of 
electrons in a single QD\cite{fuj,han1} as the Hamiltonian (\ref{eq_hf}) can be used to describe the dynamics. Consequently, the same predictions as those for a double QD can be made\cite{cak}. 
Now we derive the Hamiltonian for an electron pair in a single QD. Under a sufficiently strong magnetic field,
the triplet states $T_{\pm}$ with the magnetic quantum number of $\pm 1$ are well separated from the triplet $T_0$ state and the singlet state $S$. Thus the 
Hamiltonian within the subspace spanned by $T_0$ and $S$ states will be considered. 
The wavefunctions for the $S$ and $T_0$ states are given, respectively, as
\begin{eqnarray}
&&\Psi_S({\bf r}_1, \xi_1, {\bf r}_2, \xi_2)=\phi_g({\bf r}_1) \phi_g({\bf r}_2) 
\frac{1}{\sqrt{2}}(\alpha(\xi_1)\beta(\xi_2)-\beta(\xi_1)\alpha(\xi_2)) \;, \\
&&\Psi_{T_0}({\bf r}_1, \xi_1, {\bf r}_2, \xi_2)=\frac{1}{2}(\phi_g({\bf r}_1) \phi_e({\bf r}_2)
-\phi_e({\bf r}_1) \phi_g({\bf r}_2)) \nonumber \\
&& \hspace{3cm} \cdot (\alpha(\xi_1)\beta(\xi_2)+\beta(\xi_1)\alpha(\xi_2)) \;,
\end{eqnarray}
where $\phi_g\;(\phi_e)$ is the ground (excited) orbital state in the QD and $\alpha\;(\beta)$ denotes the spin
up (down) state. The HF interaction for two electrons is given by
\begin{equation}
V_{HF}=A v_0 \sum_i {\bf S}_1 \cdot {\bf I}_i \; \delta({\bf r}_1-{\bf r}_i) 
+ A v_0 \sum_i {\bf S}_2 \cdot {\bf I}_i \; \delta({\bf r}_2-{\bf r}_i) \;,
\end{equation}
where $A$ is the HF coupling constant, $v_0$ the volume of a unit cell, {\bf S} ({\bf I}) is the 
electron (nuclear) spin operator and the summation is carried out over nuclear spins at the location 
${\bf r}_i$. Then we find
\begin{eqnarray}
&&\langle \Psi_{T_0}|V_{HF}|\Psi_{S}\rangle= -\frac{1}{\sqrt{2}} A v_0 \sum_i \phi_e^*({\bf r}_i)
\phi_g({\bf r}_i) I_{iz} \;,  \\
&& \langle \Psi_{S}|V_{HF}|\Psi_{T_0} \rangle=
\langle \Psi_{T_0}|V_{HF}|\Psi_{S} \rangle^* \;, \langle \Psi_{S}|V_{HF}|\Psi_{S}\rangle = \langle \Psi_{T_0}|V_{HF}|\Psi_{T_0}\rangle =0\;.
\end{eqnarray}
Thus the singlet-triplet mixing is induced by the HF interaction. The effective nuclear
field operator will be denoted by
\begin{equation}
h= -\frac{1}{\sqrt{2}} A v_0 \sum_i \phi_e^*({\bf r}_i) \phi_g({\bf r}_i) I_{iz}
\end{equation}
which has the dimension of energy and its mean square value is estimated as
\begin{eqnarray}
&&\langle h h^{\dag}\rangle=\frac{(A v_0)^2}{2}  \sum_i |\phi_e^*({\bf r}_i) \phi_g({\bf r}_i)|^2 
\langle I^2_{iz} \rangle \\
&&\hspace{0.95cm} =\frac{A^2 v_0}{2} \frac{I(I+1)}{3} \int d^3 r \; |\phi_e^*({\bf r}) \phi_g({\bf r})|^2 \;,
\end{eqnarray}
where $I$ is the magnitude of the nuclear spin.
Employing the envelope functions for the ground and excited states given by
\begin{eqnarray}
&&\phi_g(r, \theta, z)=\frac{1}{\sqrt{\pi}r_0} e^{-\frac{r^2}{2r_0^2}} \sqrt{\frac{2}{d}} \cos(\frac{\pi z}{d})\;, \\
&&\phi_e(r, \theta, z)=\frac{1}{\sqrt{\pi}r_0^2} e^{-\frac{r^2}{2r_0^2}} \; r e^{-i \theta} \sqrt{\frac{2}{d}} 
\cos(\frac{\pi z}{d})\;, \\
&& r_0=\sqrt{\frac{\hbar}{m^* \Omega}} \;, \; \Omega=\sqrt{\omega^2_0+(\frac{eB}{2m^* c})^2} \;,
\end{eqnarray}
where $\omega_0$ is the frequency of the harmonic confinement in the lateral direction, $m^*$ the electron effective mass and $d$ is the thickness of the QD, we have
\begin{equation}
\langle h h^{\dag}\rangle=\frac{A^2 v_0}{16 \pi r_0^2 d} \; I(I+1)\;.
\end{equation}
In the vicinity of the singlet/triplet crossing point, the relevant Hamiltonian is given by
\begin{equation}
H=\left(\begin{array}{cc}
|S\rangle & |T_0\rangle \\
0 & h^{\dag} \\
h & 0 \end{array} \right) \;, \quad h=-\frac{1}{\sqrt{2}} A v_0 \sum_i \phi_e^*(\vec{r}_i)
\phi_g(\vec{r}_i) I_{iz} \;.
\end{equation}
Thus the same predictions of the bunching and revival phenomena as before can be made since the relevant Hamiltonian is the same.

\section{SUMMARY}

We have investigated theoretically the fundamental elements for realizing the quantum repeater 
based on photons as flying qubits and electrons as operation qubits in semiconductor nanostructures; 
namely, the quantum state transfer between
a photon and an electron spin, the quantum correlation (Bell) measurement of two electrons and the 
electron-nuclei coupled dynamics whose understanding is indispensable to realize the nuclear spin 
quantum memory.
For the first element, we have analyzed the performance of the quantum state transfer
and clarified the conditions to be satisfied to achieve a high value of the purity of the transferred quantum state or the 
fidelity of the operation. For the second element, we proposed an optical method to distinguish between
four states of two electrons based on the Faraday or Kerr rotation and confirmed the feasibility.
For the third element, we studied the electron-nuclei coupled dynamics and predicted a couple of new phenomena related
to the correlations induced by the hyperfine interactions.
The underlying mechanism is the squeezing or the increase in the purity of
the nuclear spin state through the electron spin measurements. 
 We can construct hopefully a secure and robust system of the quantum repeater combining these results,
namely, the efficient quantum state transfer between a photon and an electron spin, the reliable Bell 
measurement of two electrons for the entanglement swapping based on the Faraday or Kerr rotation and 
the long-lived quantum memory based on nuclear spins. 

\acknowledgments

 We thank the financial supports from the CREST project of 
the Japan Science and Technology Agency, from the SCOPE program of the
Ministry of Internal Affairs and Communications, and from the Ministry of Education, 
Culture, Sports, Science and Technology, Japan. The numerical calculation in this work has been done using the facilities of the Supercomputer Center, Institute for Solid State Physics, University of Tokyo.


\end{document}